\documentclass{sadhana}

\usepackage{graphicx,amsmath,bm}
\usepackage{multirow}
\usepackage{algorithm,algorithmic}

\newcommand{\algorithmicbreak}{\textbf{break}}
\newcommand{\BREAK}{\STATE \algorithmicbreak}

\begin{document}

\title{Innovative ranking strategy for IPL team formation}


\author{Saptarshi Banerjee\textsuperscript{1}, Arnabi Mitra\textsuperscript{1}, Debayan Ganguly\textsuperscript{1}, Ritajit Majumdar\textsuperscript{2} \and Kingshuk Chatterjee\textsuperscript{3}}
\affilOne{\textsuperscript{1} Department of Computer Science and Engineering, Government College of Engineering and Leather Technology\\}
\affilTwo{\textsuperscript{2} Advanced Computing \& Microelectronics Unit, Indian Statistical Institute, Kolkata\\}
\affilThree{\textsuperscript{3} Department of Computer Science and Engineering, Government College of Engineering and\\ Ceramic Technology\\
sjee712702@gmail.com, arnabimitra100@gmail.com, debayan3737@gmail.com,\\ kingshukchaterjee@gmail.com, majumdar.ritajit@gmail.com}


\twocolumn[{

\maketitle

\begin{abstract}
 Indian Premier League (IPL) is a tournament of twenty over cricket matches. Teams of this tournament are selected via an auction from a pool of players. Each team employs a think-tank to build the best possible team. Few studies have been performed to automate the process of team selection. However, those studies mostly concentrate either on the current form of the players, or their long term performance. In this paper, we have (i) selected traditional features as well as determined some derived features, which are generated from the traditional features, for batsmen and bowlers, (ii) formulated heuristics for clustering batsmen into openers, middle order batsmen and finishers, (iii) formulated heuristics for relative ranking of batsmen and bowlers considering the current performance as well as the experience of each player, and (iv) have proposed two greedy algorithms for team selection where the total credit point of the team and the number of players in each cluster is fixed. Our proposed ranking scheme and algorithm not only determines the best possible team, but can also determine the best alternate player if one of the target players is unavailable.
\end{abstract}



}]




\markboth{S. Banerjee et al.}{IPL Team Formation}

\section{Introduction}
Cricket is the most popular sports in India. Different format of this game has gained popularity in different times, and in recent times Twenty-twenty (T20) format of the game has gained popularity. In last eleven years, Indian Premier League (IPL) has created a position of its own in the world cricket community. In this tournament, there are eight teams named after eight cities of India. Each team is owned by one or more franchises. A pool of players is created for an auction. Each player is allotted a base price, and the maximum amount each franchise can spend for its entire team is fixed. The auction determines the team for each franchise.

Naturally, the aim of every franchise is to buy the best players from the auction who can help them win the tournament. Each franchise, therefore, maintains a think-tank whose primary job is to determine the players whom they want to buy from the auction. This is not a trivial task because (i) it is not possible to always pick the best players since the budget for a team is fixed, and (ii) often it so happens that some other franchise buys a player who was in the target list of a franchise. In such situations, the think-tank must determine the best alternate player for the team. The history of IPL has repeatedly seen teams failing to perform in the tournament due to poor player selection.

In this paper, we have developed a recommendation system for player selection based on heuristic ranking of players, and a greedy algorithm for the team selection. The algorithm can help the think-tank to determine the best potential team, and an alternate player if their target player is not available. Previous studies \cite{dey, prakash, shah, sharma} concentrate either on the current form of the players \cite{Carl, Moore}, or their long term performance history \cite{Harsha}. However, these two factors individually are not sufficient to decide whether a player is to be bought. Other factors such as, whether a batsman is an opener, or middle-order player or finisher, and consideration of features pertinent to those ordering is of utmost necessity. For example, a middle-order batsman can afford a lower strike rate if he has a good average, but not a finisher. Furthermore, it is necessary to determine a balance between the recent form of a player, and the past history of a good player whose recent form may not be up to the mark.

In this paper, we have considered a set of traditional and derived features and have quantified them. Not all of these features are equally important for every player in every position. Therefore, we have broadly classified a potential team into multiple positions, and for each position we have heuristically determined the appropriate weight for these features. For each player in the pool, we have obtained a score based on these weighted features, and have ranked them accordingly. The ranking obtained by this technique is in accordance with the well known ranking of players in IPL. Finally, a relative score on the scale of 1-10 is allotted for each player. Moreover, a fixed basis score is allocated for a team of 15 players, which emulates the fixed budget assigned to each team. We then use greedy algorithm to select the best team within this budget using the aforementioned ranking scheme.

The rest of the paper is organized as follows - In Section 2, we define the traditional and derived features which have been considered for batsmen, and quantify them. Three clusters - openers, middle-order and finishers, have been defined in Section 3, each having a heuristic scoring formula which is a weighted sum of those features. The batsmen have been ranked into these clusters according to their points by these heuristics. The features for the bowlers are quantified in Section 4 and the bowlers are ranked accordingly. In Section 5, we further assign credit points to the players according to their ranks. We present two greedy algorithms for selecting the best IPL team from the previous ranking when the total credit point of the team is fixed. We conclude in Section 6.

\section{Analyzing features for batsmen}

We have created a database of all the players and their performance in the last eleven seasons of IPL. Some players, who have already retired, are removed from the database. The performance values for those players who have not played some of the early seasons are assigned 0 for those seasons. This comes handy later on while determining the experience factor. For analysis of current form, we have considered the values from the 2018 season of IPL only. In Table~\ref{tab:traditional} we note the traditional features which are considered for the analysis of players. These are very standard features used to report the performance of players in every cricket matches \cite{david}, and hence we do not discuss about these. Apart from these features, some derived features are also quantified, which we shall discuss later in this section.

\begin{table}[htb]
    \centering
    \caption{Standard features for batsmen and bowlers}
    \begin{tabular}{|c|c|}
    \hline
       Batsmen & Bowlers \\
       \hline
        Innings & Innings\\
        Runs Scored & Wickets Taken\\
        \# Balls Faced & \# Balls Bowled\\
        Average & Average\\
        Strike Rate & Strike Rate\\
        \# 100s & \# Runs Conceded\\
        \# 50s & Economy Rate\\
        \# 4s Hit & \# 4 Wickets\\
        \# 6s Hit & \# 5 Wickets\\
        \hline
    \end{tabular}
    \label{tab:traditional}
\end{table}

We have grouped the batsmen into three position clusters - openers, middle order batsmen and finisher, since these three types of players have three very different role in the match. In the remaining part of this section, we have quantified the features considered in Table~\ref{tab:traditional} for grouping. However, these features conform to all batsmen, and hence are not sufficient for the clustering. Therefore, we also consider some derived features which take into account the specific roles of batsmen in different position clusters.

\begin{itemize}
    \item \textbf{Batting Average}: This feature denotes the average run scored by a batsman per match before getting out.
    \begin{center}
        Avg = Runs/(\# Innings - \# Not Out)
    \end{center}
    
    \item \textbf{Strike Rate}: Strike rate is defined as the average runs scored by a batsman per 100 balls. The higher the strike rate, the more effective a batsman is at scoring runs quickly.
    \begin{center}
        SR = ($100 \times$ Runs)/(\# Balls)
    \end{center}
    
    \item \textbf{Running Between the Wicket}: Though this is a very frequently used term in cricket, there is no proper quantification of this feature. We have quantified it as the number of runs scored per ball in which fours or sixes were not hit.
    \begin{center}
        RunWicket = (Runs - \# Fours$\times 4$ - \# Sixes$\times 6$)\\ /(\# Balls - \# Fours - \# Sixes)
    \end{center}
    
    \item \textbf{Hard Hitting}: T20 is a game of runs, and to win it is necessary to score runs quickly. Therefore, apart from quick running, it is necessary to hit many fours and sixes. ``Hard Hitter" is a common term in T20 cricket, but it is not quantified. We have quantified this feature as the number of runs scored per ball by hitting four or six.
    \begin{center}
        HardHitting = (\# Fours$\times4$ + \# Sixes$\times6$)/\# Balls
    \end{center}
\end{itemize}

In addition to these, we have defined a \texttt{cost} feature for each of the features. The set of \texttt{cost} features is used to obtain a relative score of an individual with respect to all the IPL players. Let $f(i)$ denote the value of a feature $f$ for the $i$-th player. If the total number of IPL players is $n$, then the cost feature for $f$ is defined as $f(i)$/($\max\limits_{1\le j \le n}$ \{$f(j)$\}). Using this formula, we have calculated the cost feature for each of the features discussed above.

Experience of a player is an important criteria which should be considered in addition to the above features. Therefore we have defined experience factor ($x$fact) as
\begin{center}
    $x$fact(i) = innings(i)/(\# innings in IPL so far)
\end{center}

where innings(i) implies the number of innings the $i$-th player has played. Define range$_x$fact as follows
\begin{center}
    range$_x$fact = $\max\limits_{1\le j \le n}$ \{$x$fact(j)\} - $\min\limits_{1\le k \le n}$ \{$x$fact(k)\}
\end{center}

Then the relative experience of a player (cost$_x$fact) is defined as
\begin{center}
    cost$_x$fact(i) = $x$fact(i)/range$_x$fact
\end{center}

The calculation of cost feature and cost$_x$fact is similar for bowlers also.

\section{Clustering and ranking of batsmen}
We have clustered the batsmen into three major categories - (i) opener, (ii) middle order and (iii) finisher. These three types of batsmen are required to play different roles in the match, and hence are expected to have different skills. A total weight of $100$ is divided into the features for each batsman. The division of the total weight into features is heuristic so that it models the skill requirements for batsman in different clusters. Furthermore, the ranking of players obtained by such weight distribution conforms with our known player ranking. In the following subsections we discuss the motivations for weight division in each position cluster, and show the top five players according to our ranking scheme.

\subsection{Opening batsman}
The responsibility of setting up a good foundation for the team's score lies on the openers. The openers get to face the maximum number of balls, and therefore is expected to have a high average. Furthermore, they need to score quickly in the first power play. So a handy strike rate is also a good indicator of the effectiveness of an opening batsman. Both these features are equally important and are, therefore, assigned the highest weight of 30 each. Furthermore, an opener is expected to stay on the crease for a long time and score big runs. Therefore, we have assigned a weight of 20 to the number of half-centuries (hc) scored by an opener per innings. Often an opener requires some time to set in, and then start hard hitting. During the time, when an opener is still not hitting hard, he should rotate the strikes quickly to keep the scoreboard moving. However, the necessity of hard hitting cannot be totally ignored during the powerplay. This motivates us to assign a weight of 10 for both running between the wickets and hard hitting.

Based on the choice of feature and weight division, the relative score of the $i$-th opener (opener(i)) is determined as
\begin{center}
    opener(i) = cost\_SR(i)$\times 30$ + cost\_Avg(i)$\times 30$ + (hc(i)/innings(i))$\times 20$ + cost\_RunWicket(i)$\times 10$ + cost\_HardHitting(i)$\times 10$
\end{center}

We have used the notation $f(i)$ to denote the value of the feature $f$ for the $i$-th player considering all the seasons of IPL. Another notation $f[i]$ is used to denote the value of the same feature considering only the last season of IPL. The relative current score of the $i$-th opener(curr\_opener[i]) is determined as
\begin{center}
    curr\_opener[i] = cost\_SR[i]$\times 30$ + cost\_Avg[i]$\times 30$ + (hc[i]/innings[i])$\times 20$ + cost\_RunWicket[i]$\times 10$ + cost\_HardHitting[i]$\times 10$
\end{center}

Considering the experience factor for each player, the final rank of the $i$-th opener is calculated as
\begin{center}
    opener\_rank(i) = opener(i)$\times$ cost$_x$fact$\times$ (curr\_opener[i]/mean\_opener) + curr\_opener[i]
\end{center}
where mean\_opener is the average score of all the openers.

The top five opening batsman from IPL pool of players and their corresponding point derived according to our ranking scheme is shown in Table~\ref{tab:Open}.

\begin{table}[htb]
    \centering
    \caption{Top five opening batsmen according to our ranking scheme}
    \begin{tabular}{|c|c|}
    \hline
        Batsman & Points \\
        \hline
        AB de Villiers & 173.5798\\
        MS Dhoni & 159.0942\\
        DA Warner & 150.113\\
        V Kohli & 133.8061\\
        CH Gayle & 132.4749\\
        \hline
    \end{tabular}
    \label{tab:Open}
\end{table}

Four out of the five names are indeed the top openers or first down batsmen in IPL. The striking inclusion in this table is MS Dhoni who is almost always a finisher. However, we shall see in the subsequent subsections that the points obtained by Dhoni as a finisher is significantly higher than his points as an opener. That his name appeared in this table simply shows the effectiveness of Dhoni in a T20 match.

\subsection{Middle order batsman}
The batsmen in these genre need to provide the stability and also must possess the ability to accelerate the scoreboard when chasing a big total. A middle order batsman must be a good runner between the wickets since it becomes difficult to hit big shots during this phase of the match with the fielders spread out. Furthermore, often when one or both the openers get out quickly, the middle order batsmen must take up to responsibility to score big runs. Therefore a decent average is necessary.

The weights for middle order batsmen have been distributed among the features taking the above requirements into consideration. The relative score of the $i$-th middle order batsman (middle(i)) is determined as
\begin{center}
    middle(i) = cost\_SR(i)$\times 20$ + cost\_Avg(i)$\times 30$ + (hc(i)/innings(i))$\times 10$ + cost\_RunWicket(i)$\times 25$ + cost\_HardHitting(i)$\times 15$
\end{center}

In accordance with the calculation for openers, the relative current score of the $i$-th middle order batsman (curr\_middle[i]) is determined as
\begin{center}
    curr\_middle[i] = cost\_SR[i]$\times 20$ + cost\_Avg[i]$\times 30$ + (hc[i]/innings[i])$\times 10$ + cost\_RunWicket[i]$\times 25$ + cost\_HardHitting[i]$\times 15$
\end{center}

Considering the experience factor for each player, the final rank of the $i$-th middle order batsman is calculated as
\begin{center}
    middle\_rank(i) = middle(i)$\times$ cost$_x$fact$\times$ (curr\_middle[i]/mean\_middle) + curr\_middle[i]
\end{center}

where mean\_middle is the average score of all the middle order batsmen.

Based on the middle\_rank, we have sorted all the batsmen in descending order of their score. The top five middle order batsmen, according to our scoring scheme is shown in Table~\ref{tab:middle}.

\begin{table}[htb]
    \centering
    \caption{Top five middle order batsmen according to our ranking scheme}
    \begin{tabular}{|c|c|}
    \hline
    Batsman & Points\\
    \hline
        AB de Villiers & 183.9566 \\
        MS Dhoni & 169.7258\\
        DA Warner & 163.6285\\
        V Kohli & 150.5608\\
        KD Karthik & 137.0331\\
        \hline
    \end{tabular}
    \label{tab:middle}
\end{table}

Once again, the names in this ranking do not require any justification. It is worthwhile to note that Dhoni is present in this list also, and his score is slightly higher than his score as an opener. This shows that Dhoni is more effective as a middle order batsman.

\subsection{Finisher}
Finishers usually have the task of scoring quick runs in the end of the match. Naturally, strike rate and hard hitting are the most important factors for any finisher. It is difficult for a finisher to score big runs regularly since they usually get to play very few overs. Therefore, average score is not considered for these players. Running between the wicket is also an important factor for these batsmen. These players are also expected to remain not out and win the match for the team.\\

In accordance to the above requirements, we have calculated the relative score of the $i$-th finisher (finisher(i)) as follows
\begin{center}
    finisher(i) = cost\_SR(i)$\times 40$ + cost\_HardHitting(i)$\times 40$ + not\_out(i)$\times 5$ + cost\_RunWicket(i)$\times 15$
\end{center}

The current form of the $i$-th finisher (curr\_finisher) is calculated considering only the feature scores for last year.
\begin{center}
    cur\_finisher[i] = cost\_SR[i]$\times 40$ + cost\_HardHitting[i]$\times 40$ + not\_out[i]$\times 5$ + cost\_RunWicket[i]$\times 15$
\end{center}

Mean\_finisher is the average score of all the middle order batsmen. Eventually the total score of the $i$-th finisher, considering the experience factor is calculated as follows.
\begin{center}
    finisher\_rank(i) = finisher(i)$\times$ cost$_x$fact$\times$ (curr\_finisher[i]/mean\_finisher) + curr\_finisher[i]
\end{center}

Based on the score of finisher\_rank, the top five finishers in IPL are showed in Table~\ref{tab:finish} which clearly shows that Dhoni should be used as a finisher rather than an opener or middle-order batsman.

\begin{table}[htb]
    \centering
    \caption{Top five finishers according to our ranking scheme}
    \begin{tabular}{|c|c|}
    \hline
    Batsman & Points\\
    \hline
        MS Dhoni & 364.3758 \\
        DJ Bravo & 248.9014\\
        AB de Villiers & 223.4076\\
        YK Pathan & 215.7580\\
        KD Karthik & 214.2518\\
        \hline
    \end{tabular}
    \label{tab:finish}
\end{table}

Having obtained the score for each player in these three categories, we assign one or more labels (O (Opener), M (Middle Order), F (Finisher)) to the players. The category in which the player has the maximum score is naturally assigned as a label for that player. However, if a player has a higher (or equal) rank in some other category, then that category is also assigned to that player. Such players can be used interchangeably among those categories. For example, Dhoni is assigned only as a finisher since both his rank and his score is higher as a finisher than the other two categories. However, de Villiers has a higher score as a finisher, but a better rank as a middle order or opening batsman. So he can be used interchangeably among these three categories. Similarly, Karthik can be used both as a middle order batsman or as a finisher.

\section{Analyzing features for bowlers}

Similar to batsmen, we have considered a set of parameters for bowlers and have quantified them. The features which have been considered are as follows -

\begin{itemize}
    \item \textbf{Wicket Per Ball}: It is defined as the number of wickets taken per ball.
    \begin{center}
        wicket\_per\_ball = (\# wickets taken)/(\# balls)
    \end{center}
    
    \item \textbf{Average}: It denotes the number of runs conceded per wicket taken.
    \begin{center}
        Ave = (\# runs conceded)/(\# wickets taken)
    \end{center}
    
    \item \textbf{Economy rate}: Economy rate for a bowler is defined as the number of runs conceded per over bowled.
    \begin{center}
        Eco = (\# runs conceded)/(\# overs bowled) = (\# runs conceded * 6)/(\# balls)
    \end{center}
\end{itemize}

We have not clustered the bowlers into groups. Instead we have considered two parameters for a good bowler into the same heuristic. A bowler who can take 4 or 5 wickets should be included in the team. However, it is better to take a bolwer who can take 1 or 2 wickets per match rather than a bolwer who takes 4 or 5 wickets once in a while. Strike rate and consistency has been together quantified for the $i$-th bowler as
\begin{center}
    bowler(i) = (4*four(i) + 5*five(i) + wicket(i))*6/ball(i)
\end{center}

where four(i) and five(i) denote the number of matches where the bowler took 4 and 5 wickets respectively, whereas wicket(i) denote the total number of wickets taken in those matches where the bowler did not take 4 or 5 wickets. Taking the other features into consideration, we divide a total weight of 100 as follows -
\begin{center}
    bowler\_val(i) = wicket\_per\_ball(i)*35 + bowler(i)*35 + (1/Ave(i))*10 + (1/Eco(i))*10
\end{center}

The current form of a bowler (curr\_bowler\_val) is calculated similarly considering only the last season's values. The total point of a bowler, considering both current and overall form, is denoted as
\begin{center}
    final\_bolwer(i) = bowler\_val(i) $\times$ (curr\_bowler\_val(i)/mean\_bowler) $\times$ cost$_x$fact + curr\_bowler\_val(i)
\end{center}

Based on this ranking scheme, we show the top 5 bowlers in Table~\ref{tab:bowl}.

\begin{table}[htb]
    \centering
    \caption{Top five bowlers according to our ranking scheme}
    \begin{tabular}{|c|c|}
    \hline
    Bowler & Points\\
    \hline
        A. Tye & 335.3772 \\
        A. Mishra & 296.390\\
        S. Narine & 254.321\\
        P. Chawala & 223.7809\\
        R. Jadeja & 223.283\\
        \hline
    \end{tabular}
    \label{tab:bowl}
\end{table}

In the next section we provide the algorithms for selecting a team of 15 players, where the budget is fixed.

\section{Greedy algorithm for team selection}

In this section, we propose two greedy algorithms for team selection. Each team, containing $n$ players, is partitioned into the following buckets: $B$ = \{Opener, Middle-order, Finisher, Bowler\}, where each bucket $B[i]$ is a set of $k_i$ players such that
\begin{equation}
\label{eq:com}
    \sum_{i \in B} k_i = n
\end{equation}
Each team is allotted a \emph{value}, which emulates the total budget for a team. The number of players $k_i$ in each bucket $B[i]$ is decided by the user, and the \emph{unit} for each bucket is $unit(B_i) = (value*k_i)/4$.

\subsection{Assigning credit points to players}
Players in each cluster are further assigned credit points based on their ranking. This helps us to emulate the base price of a player. If a cluster contains $c_n$ players, and it is partitioned into $c_p$ credit point groups, then each group contains $c_n$/$c_p$ players. The first $c_p$ players are assigned a to the highest credit point group, the next $c_p$ players to the second highest credit point group and so on. Finally each group is assigned a credit point that decreases as we go down the groups. This step is necessary because the fixed budget of each team has been emulated as a fixed value for each team. The total credit of the team should not exceed the fixed value.

In the example of the following subsection, we have considered four credit groups with valuation 10,9,8 and 7. However, the number of groups, as well as the valuation can be varied according to the team selection criteria.

\subsection{First greedy algorithm}
In our first algorithm, we consider wicket-keepers as a separate bucket. Therefore, for our first algorithm, Equation~\ref{eq:com} is modified as $\sum_{i \in B} k_i + w = n$, where $w$ is the number of wicketkeepers. In Algorithm~\ref{greedy1}, we show our first algorithm for team selection.

\begin{algorithm}[htb]
\caption{Greedy Algorithm 1 for team selection}
\begin{algorithmic}[1]
\REQUIRE The pool of players clustered into one or more of the buckets \emph{opener, middle-order, finisher} and \emph{bowler} along with their corresponding rank. The total number of players $n$ in a team, the total valuation \emph{value} of the team, the number of players $k_i$ in each bucket $B[i]$ and the number of wicketkeepers $w$ in the team, such that $\sum_{i \in B} k_i + w = n$.
\ENSURE An optimal team of $n$ players.
\STATE unit $\leftarrow$ value/5
\FORALL{$b \in$ \{Wicketkeeper, Opener, Middle-order, Finisher, Bowler\}}
\STATE $cap_b$ $\leftarrow$ $unit \times k_b$
\STATE $min_b$ $\leftarrow$ minimum credit point of a player in bucket $b$
\STATE rem $\leftarrow$ $cap_b$
\FOR{\emph{pos} in 1 to $k_b$}
\IF{rem $< min_b$}
\WHILE{True}
\STATE j $\leftarrow$ pos
\WHILE{rem $< min_b$}
\STATE j = j-1
\IF{(credit at j)-1 $\geq min_b$}
\STATE credit at j = (credit at j)-1
\STATE rem = rem+1
\ENDIF
\IF{rem $\geq min_b$}
\BREAK
\ENDIF
\ENDWHILE
\IF{rem $\geq min_b$}
\BREAK
\ENDIF
\ENDWHILE
\ENDIF
\STATE Assign the highest credit $\leq$ rem in pos
\STATE rem = rem - assigned credit
\ENDFOR
\ENDFOR
\end{algorithmic}
\label{greedy1}
\end{algorithm}

Algorithm~\ref{greedy1} assigns the best possible credit for each player position. The total credit of each position is bounded by the value of \emph{unit}. If the algorithm comes across any position where the remaining unit is less than the minimum credit in the player pool, then it backtracks and reduces the credits assigned in the previous positions till a player is assignable in the current position. This, being a greedy algorithm, has the risk that it may end up assigning a few players with best rankings along with a few players with very low ranking.

We now produce a team of 15 players using the proposed algorithm. If a total value of $150$ or more is assigned to the team, then players of credit 10 can be selected for each position. Furthermore, a very low value can lead to a very poor team. For our example, we have chosen a value of 135, such that the unit is 9. Furthermore, in our example team, we shall have two wicketkeepers, two openers, three middle-order batsman, two finishers and six bowlers.

Since, the capacity of wicketkeeper is 2, and unit is 9, a total credit of 18 can be assigned to the two wicketkeepers. We first assign a point of 10 to the first position, and the remaining 8 points is assigned to the second position. From the rank of players, who are also wicketkeepers, Dhoni is assigned in the first position with 10 points, and S. Samson is assigned to the second position with 8 points. The team of 15 players, as obtained using the Algorithm~\ref{greedy1} is shown in Table~\ref{tab:team1}.

\begin{table}[htb]
    \centering
    \caption{A team of 15 players, with a total credit point of 135, selected using Algorithm~\ref{greedy1}}
    \begin{tabular}{|c|c|c|}
    \hline
    Position & Player & Credit Point\\
    \hline
        \multirow{2}{*}{Wicketkeeper} & M.S. Dhoni & 10 \\
         & S. Samson & 8\\
        \hline
        \multirow{2}{*}{Opener} & D. Warner & 10\\
        & K.L. Rahul & 8\\
        \hline
        \multirow{3}{*}{Middle-order} & V. Kohli & 10\\
        & A.B. de Villiers & 10\\
        & F. du Plesis & 7\\
        \hline
        \multirow{2}{*}{Finisher} & D. Bravo & 10\\
        & R. Pant & 8\\
        \hline
        \multirow{6}{*}{Bowlers} & A. Tye & 10\\
        & A. Mishra & 10\\
        & T. Boult & 9\\
        & S. Al Hassan & 9\\
        & K. Jadav & 9\\
        & M. Johnson & 7\\
        \hline
    \end{tabular}
    \label{tab:team1}
\end{table}

\subsection{Second Greedy Algorithm}
In the team selected (Table~\ref{tab:team1}) using Algorithm~\ref{greedy1}, both the wicketkeepers are finishers. Therefore, the selected team ends up with four finishers. To avoid this scenario, the second greedy algorithm, which is similar to Algorithm~\ref{greedy1}, keeps an extra restriction that the two wicketkeepers should not belong to the same bucket. By keeping this restriction, the team is exactly similar to that in Table~\ref{tab:team1}, except that instead of S. Samson, we select P. Patel as the second wicketkeeper, who is an opener.

This second algorithm can be easily further modified to ensure the cluster of the wicketkeeper. For example, one can impose a restriction such as one of the wicketkeepers must be an opener. Our proposed algorithm is flexible to handle such restrictions. Moreover, using this algorithm along with the the aforementioned ranking, a franchise can easily determine the best alternate player for a position if one of their target player is not available. For example, both Warner and Gayle have credit point 10, but the rank (and point) of Warner is higher than that of Gayle. Therefore, if Warner is already selected by some other team, then he can be replaced with Gayle. If no player of credit 10 is available, then that position can be filled with players of credit 9, and so on.

\section{Conclusion}
In this paper, we have shown a heuristic method for IPL team selection. For each player, we have considered some traditional and derived features, and have quantified them. We have clustered the players into one or more of the clusters - Opener, Middle-order, Finisher and Bowler according to the score achieved from those features. We have also taken into consideration both the current from and the experience of a player for such ranking. The ranking obtained by our heuristic scheme is in acceptance with the known player rankings in IPL. Finally we have proposed two greedy algorithms to select the best possible team from this ranking when the total credit point and the number of players in each bucket is fixed.

The future scope of this paper is to incorporate two higher level clusters of batting and bowling allrounders. The selection of the team can also include some more flexible buckets where allrounders are given higher preference than batsman and bowlers. A trade-off between inclusion of an allrounder in the team or a batsman or bowler with higher credit point can be studied. Furthermore, the greedy algorithm for team selection has a shortcoming that it may select some high ranking players with some very low ranking ones. A dynamic programming approach may be studied to ensure more or less equal quality players in the team.











\end{document}